\documentclass[sigconf]{acmart}

\AtBeginDocument{%
  }

\copyrightyear{2023}
\acmYear{2023}
\setcopyright{acmlicensed}
\acmConference[SIGSPATIAL '23]{The 31st ACM International Conference on Advances in Geographic Information Systems}{November 13--16, 2023}{Hamburg, Germany}
\acmBooktitle{The 31st ACM International Conference on Advances in Geographic Information Systems (SIGSPATIAL '23), November 13--16, 2023, Hamburg, Germany}
\acmPrice{15.00}
\acmDOI{10.1145/3589132.3625597}
\acmISBN{979-8-4007-0168-9/23/11}

\begin{document}

\title{MaaSDB: Spatial Databases in the Era of Large Language Models (Vision Paper)}

\author{Jianzhong  Qi}
\affiliation{%
  \institution{The University of Melbourne}
  \country{Australia}
}
\email{jianzhong.qi@unimelb.edu.au}

\author{Zuqing Li}
\affiliation{%
  \institution{The University of Melbourne}
  \country{Australia}
}
\email{zuqingl@student.unimelb.edu.au}

\author{Egemen Tanin}
\affiliation{%
  \institution{The University of Melbourne}
  \country{Australia}
}
\email{etanin@unimelb.edu.au}

\renewcommand{\shortauthors}{Qi et al.}

\begin{abstract}
  Large language models (LLMs) are advancing rapidly. 
  Such models have demonstrated strong capabilities in learning from large-scale (unstructured) text data and answering user queries. 
 Users do not need to be experts in 
  structured query languages to interact with systems built upon such models. This provides great opportunities to reduce the barrier of information retrieval for the general public. By introducing LLMs into spatial data management, we envisage an LLM-based spatial database system to learn from both structured and unstructured spatial data. Such a system will offer seamless access to spatial knowledge for the users, thus benefiting individuals, business, and government policy makers  alike.
\end{abstract}

\ccsdesc[500]{Information systems~Spatial-temporal systems}
\ccsdesc[500]{Information systems~Database query processing}

\keywords{Spatial Databases, Large Language Models, Model as a Database}

\maketitle

\section{Introduction}

Modern  \emph{machine learning} (ML) techniques have made breakthroughs in computer vision, natural language processing, and many other applications domains, resulting in significant performance improvements. While ML studies focus on model accuracy, and database research centers on query and transaction management efficiency, ML techniques have made their way into the database community. 
Many database problems, such as as data indexing, query optimization, and knob tuning, are being solved with a new generation of approaches that have been summarized with the term ``AI4DB''~\cite{AI4DB}. 

Spatial databases make no exception. Studies have proposed ML-based spatial indices~\cite{RSMI,10.1145/3588917}, spatial query optimizers~\cite{10.1145/3557915.3560960}, and spatial data representations~\cite{DBLP:conf/edbt/ChangTC023,t3s}. These studies have focused on using ML to optimize the effectiveness or efficiency of modules of spatial database systems. \emph{The ML models developed help retrieve query results but do not change the general query processing paradigm}: (1)~Users submit queries in a special query language (e.g., SQL)\footnote{There are studies on using ML models to translate queries in natural language into SQL queries (text-to-SQL)~\cite{DBLP:journals/vldb/KatsogiannisMeimarakisK23}, but not yet for spatial queries to the best of our knowledge.}, which are parsed and translated into relational algebra expressions. (2)~A query optimizer module analyzes the expressions and determines the order of execution (i.e., computes a query plan). (3)~The expressions are executed by a query execution engine,  with the help of spatial indices to streamline the execution. (4)~Relevant results are retrieved from the data tables and are returned to the users. This process may be repeated with altered queries until the intended results are found for the users. 

The latest development of ML techniques, i.e.,  \emph{large language models} (LLMs) such as ChatGPT, offer great opportunities to develop the next-generation spatial databases where, instead of using ML models for spatial database optimization, we envisage to use \emph{machine learning models as a spatial database} (MaaSDB). 

LLMs are neural network models with billions of parameters trained on a large text corpus. 
While such models are trained on simple tasks of predicting the next word in a sentence, they can capture the syntax and semantics of human language with a high accuracy. They can generate text and engage with human users  in dialogues,  to answer user questions and to perform text processing tasks. 
Most importantly, such models have been shown to be  able to ``memorize'' the facts from the training data~\cite{10.1162/daed_a_01905}, effectively making themselves large data repositories with rich information.  

In this paper, we present the vision of the next-generation spatial database systems exploiting the capability of LLMs to memorize facts from training data. We shift ML-based spatial database optimization from ML models \emph{for} spatial databases to ML models \emph{as} spatial databases. Such systems consist of ML models trained on structured and unstructured spatial data, which can generate query answers directly instead of retrieving data from tables.  

There are important advantages that come with such systems:

(1)~The ML models in such systems can learn from both structured and unstructured spatial data (e.g., tables and free text) and generate query results based on both types of data, unlike traditional systems where typically only either type of data are available, and it is difficult to link both types of data to answer  complex queries. 

(2) Since the ML models have full knowledge about the spatial data in the database, their inbuilt natural language-based user interface will be able to  understand user intent better than existing text-to-SQL systems do, which have  limited information about the underlying data. This leads to more relevant results returned from such systems and hence higher system usability. Such systems will significantly enhance the accessibility of spatial knowledge entailed in data stored in spatial databases. For example, a tourist may request  such systems to \emph{generate a half-day trip in Hamburg within walking distance from the conference venue of SIGSPATIAL'23}; an urban planner may request such systems to  \emph{return the top-10 suburbs with the highest electric vehicle ownership-charging station ratios}. Search engines like Google may retrieve partial answers to such queries through keyword matching, which however are limited by the availability of directly matched web documents,  again due to the retrieval nature of the query processing procedure.

We make the following contributions: 
(1) We envisage a unified spatial 
database system that uses ML models as its core query execution engine to optimize user accessibility and system usability. 
(2) We conduct a pilot experimental
study to verify the feasibility of  such a system.
(3) We
identify key research challenges and opportunities in realizing such a system.

\section{Related Work}\label{sec:related_work}
We review studies on ML-based (spatial) database optimization. 

Existing works focus on using ML techniques to optimize the effectiveness and efficiency of different modules of spatial database systems, e.g., using ML models to replace (e.g., RSMI~\cite{RSMI}) or to optimize (e.g., RLR-tree~\cite{10.1145/3588917}) the structure of traditional spatial indices~\cite{DBLP:journals/tods/0003QSH14}. A study~\cite{10.1145/3557915.3560960} trains autoencoder models to compute \emph{spatial embeddings}, i.e., vectors encoding dataset characteristics such as distribution to help predict  range query selectivity for spatial query optimization. Other studies compute embeddings for spatial objects (e.g., road segments~\cite{DBLP:conf/edbt/ChangTC023} or trajectories~\cite{t3s}) to encode their spatial features for spatial query processing. In these studies, the ML models are second-class citizens -- they help retrieve query results but do not change the classic retrieval-based query  paradigm. 

A few other studies use ML models to answer spatial queries directly. For example, Qi et al.~\cite{DBLP:conf/edbt/000100Z20} train a feedforward neural network (FFN) to predict the shortest-path distance given two points on a road network. Zeighami et al.~\cite{10.14778/3510397.3510404} train  FFNs  to predict the answer for range count queries. While these studies show that ML models can  memorize facts from spatial data, they focus on aggregate queries. Their models output scalar values and not data records. They do not have a natural language-based user interface.  

In a broader context of database research, there are text-to-SQL studies~\cite{DBLP:journals/vldb/KatsogiannisMeimarakisK23} that train ML models to translate textual queries into SQL queries, thus providing a natural language-based user interface.  These models may exploit meta data such as column names of the data tables. However, they do not generate query results directly and typically do not access the actual data records at  training. 

Motivated by the strong performance of LLMs, several vision papers~\cite{10.14778/3447689.3447706,Giovanni2023,Saeed2023} use pre-trained LLMs or  \emph{transformer} (the building block of LLMs)-based models trained on unstructured data to answer database queries. 
These papers share similar visions with ours in that they also envisage ML models to become first-class citizens in a database system. They differ from our vision in that they do not consider structured spatial data and the challenges.
Tan~\cite{Tan2023}  presents several  challenges on query processing over structured data with LLMs without envisaging a solution.  
A couple of other studies apply LLMs with structured data. Urban and Binnig~\cite{Urban2023} extract tables from a document using LLMs, while Nobari and Rafiei~\cite{Nobari2023} transform tables into a desired representation for better joinability. They do not use the learned models to generate query answers directly.  \emph{Overall, none of these studies consider the specific challenges and opportunities brought by LLMs to spatial databases. Our paper fills this gap.}  Musleh et al.~\cite{10.1145/3557915.3560972} envisage a BERT-based system for trajectory analysis, while Xue et al.~\cite{10.1145/3557915.3561026,10.1145/3488560.3498387} use language models for time series forecasting, 
exploiting the analogy between trajectories/time series and sentences.  Our study complements the studies by considering spatial data and queries beyond trajectories.

\section{Pilot Study}\label{sec:pilot}

\subsection{The Vision of the Future System}
\begin{figure}[h]
  \centering
  \includegraphics[width=\linewidth]{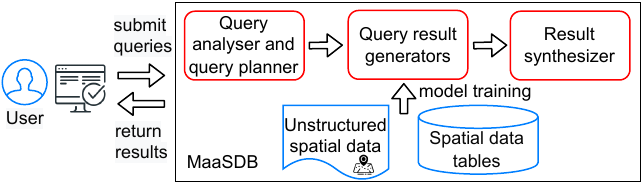}
  \caption{Overview of the future spatial database system}\label{fig:system}
\end{figure}

We envisage a next-generation spatial database system as shown 
in Fig.~\ref{fig:system}. This system consists of a \emph{query analyzer and query plan generator}, a set of \emph{query result generators}, and a \emph{result synthesizer}, which are all formed by ML models and are connected together to generate answers for user queries. The system provides a natural language-based interface for users to query the spatial knowledge learned by the ML models from spatial data stored in the system. 

Users can interact with the system 
(e.g., via a computer or a smartphone) to submit queries in natural language. Upon receiving a query (e.g., \emph{generate a half-day trip in Hamburg within walking distance from the conference venue of SIGSPATIAL'23}), the query analyzer and query plan generator (which may be an LLM) will analyze the query intent, generate sub-queries, and assign the sub-queries to the relevant subset of the query result generators (e.g., a  sub-query to find the conference venue of SIGSPATIAL'23 and a sub-query to find POIs within walking distance from the conference venue). The invoked query result  generators will generate an answer for each sub-query. Different types of query result generators will be built by training on different types of data. For example, a (transformer-based) query result generator trained on unstructured conference web pages will be able to answer the sub-query on the conference venue, while  a (GAN-based, detailed in Section~\ref{subsec:exp}) query result generator trained on a table of POIs in Hamburg will be able to answer the sub-query about the POIs.  When the results of all sub-queries have been generated, the result synthesizer (which may be another LLM) will combine them based on the user query and generate the final query answer to be returned to the user. 

Multiple challenges and research opportunities arise from the envisaged system, which will be discussed in Section~\ref{sec:conclusions}.

\subsection{Preliminary Experimental Study}\label{subsec:exp}

We verify the feasibility of the envisaged spatial database 
system through a preliminary experimental study.

\textbf{Settings.} We focus on structured data, since the aforementioned recent vision papers have shown the feasibility of using pre-trained LLMs or transformer-based models to answer certain types of database queries. We train an ML model on a  data table  and study how well the model can remember the (key characteristics of the) data. 

We use a multi-dimensional dataset (instead of a table of just spatial coordinates, for generality) named \texttt{CensusIncome}.\footnote{\url{https://archive.ics.uci.edu/dataset/20/census+income}} The dataset has 48,842 records, each with 8 categorical (e.g., occupation and marital status) and 6 numeric (e.g., age and capital gain) attributes. Since the dataset does not come with a query workload,  we follow a previous study~\cite{10.1145/153850.153878} and generate 20,000 range queries with randomly selected numeric attributes and ranges. 

We use a generative adversarial network (GAN)-based model which has been shown to be able to generate tabular data~\cite{10.14778/3231751.3231757}. 
We train a GAN model with the dataset and test how well it can generate data records that preserve the data distribution and answer range count queries, i.e., given a query range, we return the number of records in the range. A GAN model has a \emph{generator} module $G$ and a \emph{discriminator} module $D$. The 
generator generates a record given a random noise vector, $\mathbf{z}$, as its input, while the discriminator classifies whether the generated record (i.e., $G(\mathbf{z})$) resembles a real  record from the training dataset. The model is trained with a loss function that aims to generate records that cannot be distinguished from the real records.  We adapt the loss function of the generator to add a \emph{Q-Error}~\cite{10.14778/1687627.1687738} loss term, which measures how well a generated table preserves the selectivity of given range queries, as follows:
\begin{small}
\begin{equation*}\label{eq:loss}
\min_{G}\mathcal{L}(G)= \mathbb{E}_{\mathbf{z} \sim p_{\mathbf{z}}(\mathbf{z})}[\log (1-D(G(\mathbf{z})))] + \frac{1}{N}\sum_{i} \max\Big(1,\frac{sel(q_{i})}{\hat{sel}(q_{i})},\frac{\hat{sel}(q_{i})}{sel(q_{i})}\Big)
\end{equation*}
\end{small}
The second term here is the Q-Error loss, where $N$ denotes the number of range queries, $q_i$ denotes the $i$th range query, $sel(q_i)$ denotes the ground truth selectivity of 
$q_i$ on the training dataset, and $\hat{sel}(q_i)$ denotes the selectivity of $q_i$ on the generated table. We name the adapted GAN model \textbf{RC-GAN}.  
We omit the detailed model structure and hyperparameter values due to space limit. 

The experiments are run on a desktop computer with a 16-core CPU, 32 GB memory,  and 24 GB GPU memory. 

\begin{table}
  \caption{Q-Error of RC-GAN}
  \label{tab:qerror}
  \begin{tabular}{lccc}
    \toprule
    Query set &Median & 75th & 90th \\
    \midrule
    Training queries & 1.23 & 1.71 & 3.24 \\
    Testing queries & 1.25 & 1.86 & 4.08 \\
  \bottomrule
\end{tabular}
\end{table}

\textbf{Results.} We train RC-GAN (implemented with PyTorch~1.13.1) on the \texttt{CensusIncome} dataset in 10 epochs (which take about an hour) and use the trained model to generate a table of the same size of the dataset. We report the Q-Error of the generated data in Table~\ref{tab:qerror}, where ``Training queries'' refers to computing the Q-Error with the  20,000 range queries as described above, which have been used in model training, while ``Testing queries'' refers to computing the Q-Error with another set of 5,000 range queries that are generated separately (with the same procedure) and have not been seen at  training. We can see that the median Q-Errors are very close to 1 under both settings, i.e., the generated table has almost the same query selectivity as the original dataset for half of the queries. The Q-Errors at the 75th percentile are still within 2, while they only deteriorate to larger values at the 90th percentile. 
Importantly, the Q-Errors for the testing queries are close to those for the training queries. These results demonstrate the potential of ML models to ``memorize'' the key characteristics of structured data records.

We further train two Gradient Boosting classifiers with 15\% of the \texttt{CensusIncome} dataset and with 5\% of the \texttt{CensusIncome} dataset plus 10\% of data generated by RC-GAN, respectively. The classifiers predict if the income attribute of a record is greater than 50,000 given the other attributes. 
We test the classifiers on 1,000 randomly selected records of \texttt{CensusIncome} not seen at training. Table~\ref{tab:classification} reports the results. We see that the classifiers trained under both settings have very close performance, confirming the capability of RC-GAN to ``memorize'' the data distribution characteristics. 

Our results above are obtained with an ML model where the number of parameters is at the thousand scale. When larger models with more parameters are available, even better results are expected. 

\begin{table}
  \caption{Classification Accuracy with  Generated Data}
  \label{tab:classification}
  \setlength{\tabcolsep}{1mm}
  \begin{tabular}{lccc}
    \toprule
    Training data &Precision & Recall & F1 \\
    \midrule
    5\% of \texttt{CensusIncome} $+$ 10\% of RC-GAN & 0.79 & 0.97 & 0.87 \\
    15\% of \texttt{CensusIncome} & 0.82 & 0.96 & 0.88 \\
  \bottomrule
\end{tabular}
\end{table}

\textbf{Learning spatial knowledge with LLMs.} 
To provide further evidence on LLMs' potential to learn spatial knowledge, we query ChatGPT with  prompts: \emph{the geo-coordinates of the top 50 cities in Australia are} and \emph{can you give me more cities}, until 50 cities were returned. The returned geo-coordinates were correct for 49 cities, with only the geo-coordinates of Hervey Bay (a small city in Queensland) being off by 10 km. We further randomly pair up the cities to form 50 pairs. For every pair of cities $A$ and $B$, we query ChatGPT with prompt: \emph{$A$ is to which side of $B$}. The returned position results were correct for 44 pairs, with another 5 pairs obtained correct results after the geo-coordinates are further included in the prompt. Only one pair (Canberra and Orange) retained a wrong result (\emph{southwest} was returned while the answer should be \emph{south}).

These demonstrate the potential of LLMs to learn spatial knowledge and answer queries, and the research opportunities to train such models to answer more complex queries faithfully.

\section{Conclusions and Challenges}\label{sec:conclusions}

We presented a next-generation spatial database system.
This system treats ML models as first-class citizens and trains such models to ``memorize'' data stored in a spatial database and to generate query answers. It enables a new generation-based query paradigm that replaces the traditional retrieval-based paradigm. 

The system will significantly enhance the accessibility of spatial database systems, as the ML models can offer an inbuilt  natural language-based user interface and well understand users' query needs. 
It will bring huge benefits in spatial analytics and query processing, encouraging a new generation of location-based services and 
allowing better-informed location-based decision making.

To realize such a system, there are various challenges, a subset of which are summarized below.  Simply fine-tuning an open-sourced LLM such as Llama 2 directly cannot realize the system. 

(1) \emph{Faithful query result generation.} Being able to generate query results directly without an extra data retrieval process offers great opportunities to answer complex spatial (analytical) queries. 
This, however, also brings significant challenges to ensure the faithfulness of the results generated.  ML models are known to return inaccurate results. In terms of LLMs, \emph{hallucination}  is a known problem that impinges LLMs' wider applicability. When LLMs are applied to form the query engine of spatial databases, it is important to address the hallucination problem, e.g., to build a system that returns  faithfulness scores together with the generated answers. A unique opportunity arises when building such a system for spatial databases, as traditional retrieval-based query procedures can be applied in parallel to compute query answers that serve as the ground truth for training the faithfulness scoring module. 

(2) \emph{Large model training with structured spatial data.} 
There are two major issues that prevent training LLMs on structured data records directly (which are probably the reason why the other vision papers~\cite{10.14778/3447689.3447706,Giovanni2023,Saeed2023} did not take this approach): (i)~There is limited availability of structured spatial data. Comparing with the volume of free texts (e.g., web documents), the number of spatial data tables available is much smaller. The number of data tables in a spatial database is even smaller. How to train an LLM generalizable to different queries with data in such smaller scale is challenging. 
(ii)~There is incompatibility between structured spatial data and the  training procedure of LLMs.  LLMs are trained via the task of predicting the next word in a sentence. Simply treating every spatial data record as a sentence and every data field as a word to train an LLM is ineffective. This is because words in a sentence have a strong correlation, and the context of a word implies the semantics of the word. In contrast, different fields of a data record may be much less relevant, and the nearby fields of a value do not necessarily imply the semantics of the value. Further, values in a data record may be numeric and continuous, and the same value may have completely different meanings in different fields, while words are discrete and each word has much fewer different meanings. Novel model design and training procedures are needed for structured spatial data.

(3) \emph{Versatile query processing.} A problem related to the difficulty in model training given limited structured spatial data is how to answer different types of spatial queries using a model trained with limited data. While data of  limited scale may be easier to be ``memorized'' by an ML model, it does not help train a model that is generalizable to different types of queries. Also due to the limited scale of data, the trained models may not have seen too many different prompts that imply different types of queries. The generalizability of the trained models would most likely need to come from unstructured spatial data, e.g., the  Wikipedia article of a POI. Algorithms to fine-tune such models and incorporate  knowledge from structured spatial data  await exploration.

Further, the models need to have multi-step reasoning capabilities to answer complex spatial queries. For example,  to ``generate a half-day trip in Hamburg within walking distance from the conference venue of SIGSPATIAL'23'' would require (i) producing the location of the conference venue, (ii) producing POIs within walking distance around it, and (iii) selecting and ordering the POIs to form a trip. While prompting, training, and fine-tuning strategies have been proposed for this issue, achieving such advanced reasoning capabilities remains an open challenge~\cite{dziri2023faith}. 

(4) \emph{Challenges in managing ML models for data management.} There are inherent problems in data management with ML models, such as how to update the trained ML models when the underlying data have changed (e.g., moving objects~\cite{DBLP:journals/is/Wang0XQ0Y14,DBLP:journals/vldb/WardH0Q14}). Such challenges have been discussed in the literature~\cite{Giovanni2023,Saeed2023} and are not reiterated.

\vspace{-0.3mm}
\begin{acks}
This work is partially supported by Australian Research Council (ARC) Discovery Project DP230101534.
\end{acks}
\vspace{-0.3mm}
\bibliographystyle{ACM-Reference-Format}
\bibliography{ref}

\end{document}